\documentclass{cernyrep}
\usepackage[T1]{fontenc}
\usepackage[bookmarks, colorlinks=true, linktoc=page, pdftex, linkcolor=black, citecolor=black, urlcolor=blue]{hyperref}

\usepackage{fancyhdr}
\fancyhfoffset{4 mm}
\fancypagestyle{ARTTITLE}{%
\fancyhf{} 
\lhead{Proceedings of the CERN--Accelerator--School course on
{\it Introduction to Accelerator Physics}}
\lfoot{Available online at \url{https://cas.web.cern.ch/previous-schools}}
\rfoot{\thepage\hspace*{3mm}}
 
}

\newcommand{\der}[2]{\frac{\mathrm{d}#1}{\mathrm{d}#2}}
\newcommand{\pder}[2]{\frac{\partial#1}{\partial#2}}
\newcommand{\rd}[0]{\mathrm{d}}

\usepackage{varwidth}
\usepackage{xcolor}

\begin{document}

\title{Longitudinal Beam Dynamics in Circular Accelerators}
\author{F. Tecker}
\institute{CERN, Geneva, Switzerland}

\begin{abstract}
This paper gives an introduction of longitudinal beam dynamics for circular accelerators. 
After briefly discussing some types of circular accelerators, it focuses on particle motion in synchrotrons.
It summarizes the equations of motion, discusses phase-space matching during beam transfer, and introduces the Hamiltonian of longitudinal motion.
\end{abstract}
\maketitle
\thispagestyle{ARTTITLE}

\section{Introduction}
The force $\vec{F}$ on a charged particle with a charge $q$ is given by the Newton--Lorentz force:
\begin{equation}
\vec{F}=\frac{\mathrm{d}\vec{p}}{\mathrm{d}t} = q \left( \vec{E} + \vec{v} \times \vec{B} \right) \, .
\label{eq:force}
\end{equation}

The second term on the right-hand side is always perpendicular to the direction of motion, so it does not give any longitudinal acceleration and does not increase the energy of the particle.
Hence, the acceleration has to come from an electric field $\vec{E}$. To accelerate the particle, the field needs to have a component in the direction of motion of the particle. If we assume the field and the acceleration to be along the $z$ direction, Eq.~(\ref{eq:force}) becomes
\begin{equation}
\der{p}{t} = q E_z \, .
\end{equation}

The total energy $E$ of a particle is the sum of the rest energy $E_0$ and the kinetic energy $E_\mathrm{kin}$:
\begin{equation}
E = E_0 + E_\mathrm{kin} \, .
\end{equation}
In relativistic dynamics, the total energy $E$ and the momentum $p$ are linked by
\begin{equation}
E^2 = E_0^2 + p^2 c^2
\label{eq:E}
\end{equation}
(with $c$ being the speed of light), from which it follows that
\begin{equation}
\mathrm{d}E = v \, \mathrm{d}p \: .
\label{eq:de}
\end{equation}

The rate of energy gain per unit length of acceleration (along the $z$ direction) is then given by
\begin{equation}
\der{E}{z} = v \der{p}{z} = \der{p}{t} = q E_z\, , 
\end{equation}
and the (kinetic) energy gained from the field along the $z$ path follows from $ \mathrm{d}E_\mathrm{kin} = \mathrm{d}E = q E_z \,\mathrm{d}z$:
\begin{equation}
\Delta E = q \int\! E_z \,\mathrm{d}z = q \Phi \, ,
\end{equation}
where $\Phi$ is just an electric potential.

The accelerating system will depend on the evolution of the particle velocity, which depends strongly on the type of particle.
The velocity is given by
\begin{equation}
v = \beta c = c \,\sqrt{1-\frac{1}{\gamma^2}} \; ,
\end{equation}
with the relativistic gamma factor of $\gamma = E / E_0$, the total energy $E$ divided by the rest energy 
$E_0$\footnote{As there is no ambiguity with the Courant-Snyder functions in this article, I will just use $\beta$ and $\gamma$ for the relativistic beta and gamma, that might be called $\beta_r$ and $\gamma_r$ otherwise.}.
Electrons reach a constant velocity (close to the speed of light) at relatively low energies of a few mega\-electronvolts,
whereas heavy particles reach a constant velocity only at very high energies.

In particular, this requires an acceleration system that remains synchronized with the particles during their acceleration.
For instance, when the revolution frequency in a synchrotron varies, the radio frequency (rf) will have to change correspondingly.
This effect is stronger at lower particle energies (to be precise, lower relativistic~$\gamma$). 
For example, the CERN Proton Synchroton Booster was accelerating protons from a kinetic energy of $50\,$MeV to $1.4\,$GeV with a corresponding change in revolution frequency from $602\,$kHz to $1746\,$kHz, almost a factor 3.
When the protons at the Large Hadron Collider (LHC) are accelerated from $450\,$GeV to $7\,$TeV, the relative change in the revolution frequency is only $2\cdot 10^{-6}$!
As a consequence, one needs different types of accelerating structures, optimized for different velocities, with different requirements in terms of bandwidth.

\section{Phase conventions}

\begin{figure}[!h]
 \centering\includegraphics[width=0.83\textwidth]{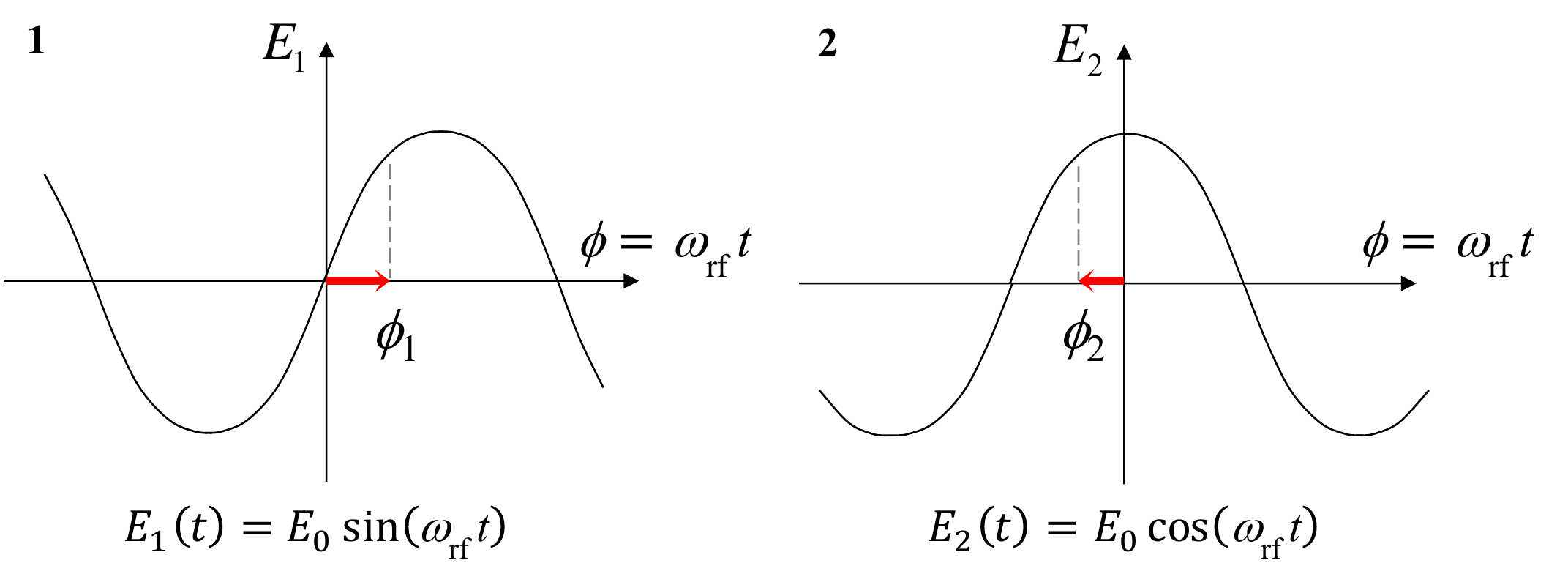}
 \caption{\label{fig:phase-conventions} Common phase conventions: (1)~the origin of time is taken at the zero-crossing with positive slope; (2)~the origin of time is taken at the positive crest of the rf voltage.}
\end{figure}

Several phase conventions exist in the literature (see Fig. \ref{fig:phase-conventions}):
\begin{itemize}
\item mainly for circular accelerators, the origin of time is taken at the zero-crossing with positive slope;
\item mainly for linear accelerators, the origin of time is taken at the positive crest of the rf voltage.
\end{itemize}
In the following, I will stick to the former convention of the positive zero-crossing, as it is more common in the circular case.

\section{Circular accelerators -- betatron}
Insulation issues limit generally the acceleration by static electric fields. Furthermore in the case of circular accelerators, 
we would not achieve any net acceleration over a full turn for a static field, since in this case
\begin{equation}
	\oint \vec{E} \cdot \rd\vec{s} = 0 .
\end{equation}
Consequently, we do need an electric field that varies with time.

In addition to the electric field generated by a scalar potential $\Phi$, the time derivative of a vector potential $\vec{A}$ also contributes to the electric field:
\begin{equation}
	\vec{E} = - \vec\nabla \Phi - \frac{\partial\vec{A}}{\partial{t}} .
\end{equation}
Since the vector potential and the magnetic field $\vec{B}$ are related by
\begin{equation}
	\vec{B} =  \mu \vec{H} = \vec\nabla \times \vec{A} .
\end{equation}
It it follows that
\begin{equation}
	\vec\nabla \times \vec{E} = -  \frac{\partial\vec{B}}{\partial{t}} ,
\end{equation}
so the time variation of the magnetic field generates an electric field, which can accelerate particles overcoming the static insulation problems and give a net acceleration in a circular accelerator.

\begin{figure}[!b]
\vspace*{4mm}
 \centering\includegraphics[width=0.53\textwidth]{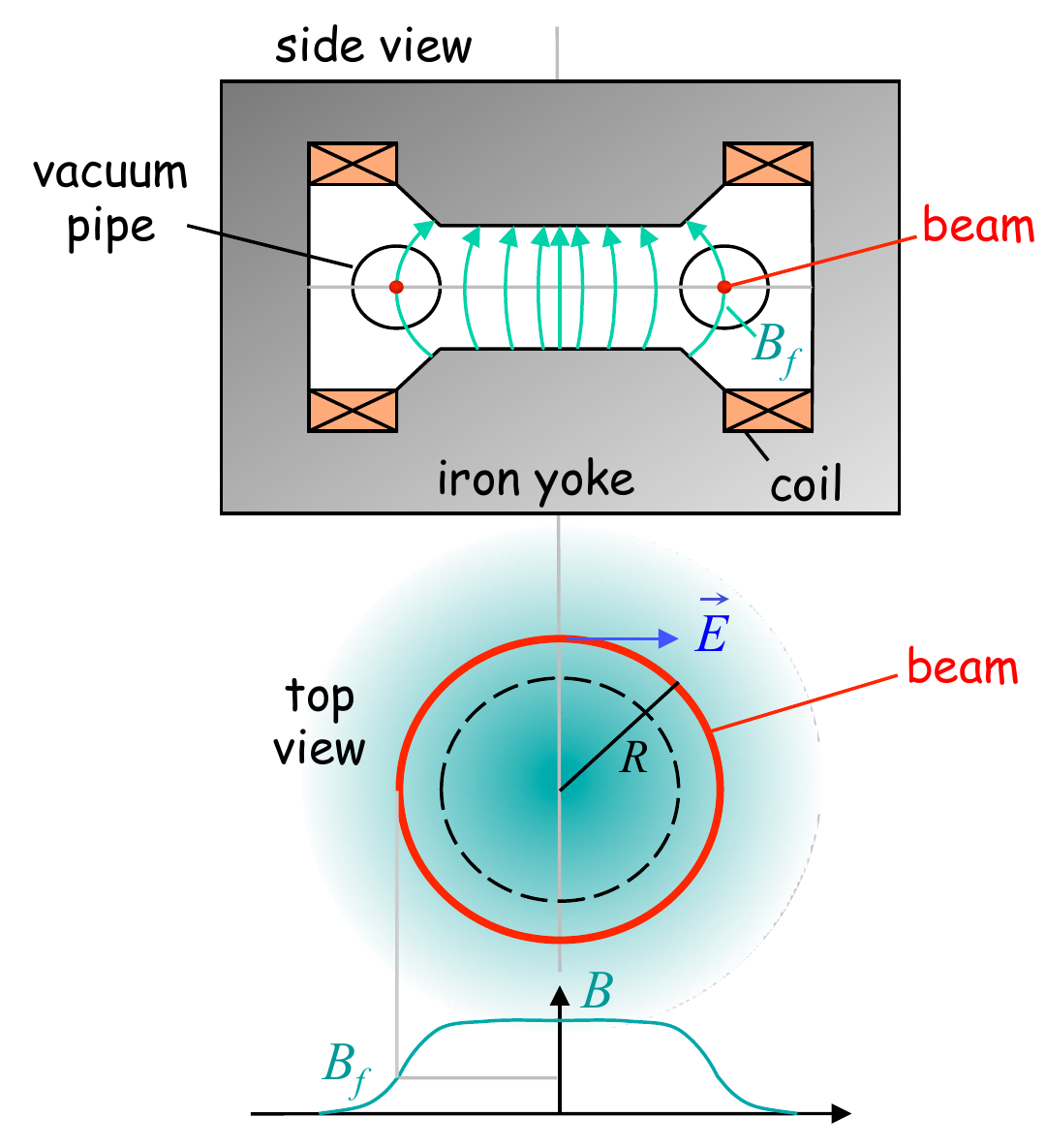}
 \caption{\label{fig:betatron} Schematic of a betatron. The top shows the side view, the middle the top view, and the graph at the bottom the magnetic field distribution.}
\end{figure}

One method based on this principle is the {\em betatron}, as shown in Fig.~\ref{fig:betatron}.
The circularly symmetric magnet is fed by an alternating current at a frequency typically between 50 and $200\,$Hz.
The time-varying magnetic field $\vec{B}$ creates an electric field $\vec{E}$ and at the same time guides the particles on a circular trajectory.
According to the field symmetry, the electric field generated is tangent to the circular orbit.
A~more detailed description can be found in~\cite{bib:CAS94-LeDuff}.

Betatrons were used to accelerate electrons up to about 300\ MeV with the energy reach limited by the saturation in the magnet yoke.
They are still in use at lower energies for X-ray sources for industrial radiography and for medical radiation therapy.

\section{Circular accelerators -- cyclotron}
\begin{figure}[!h]
\vspace*{4mm}
 \centering{\includegraphics[width=0.45\textwidth]{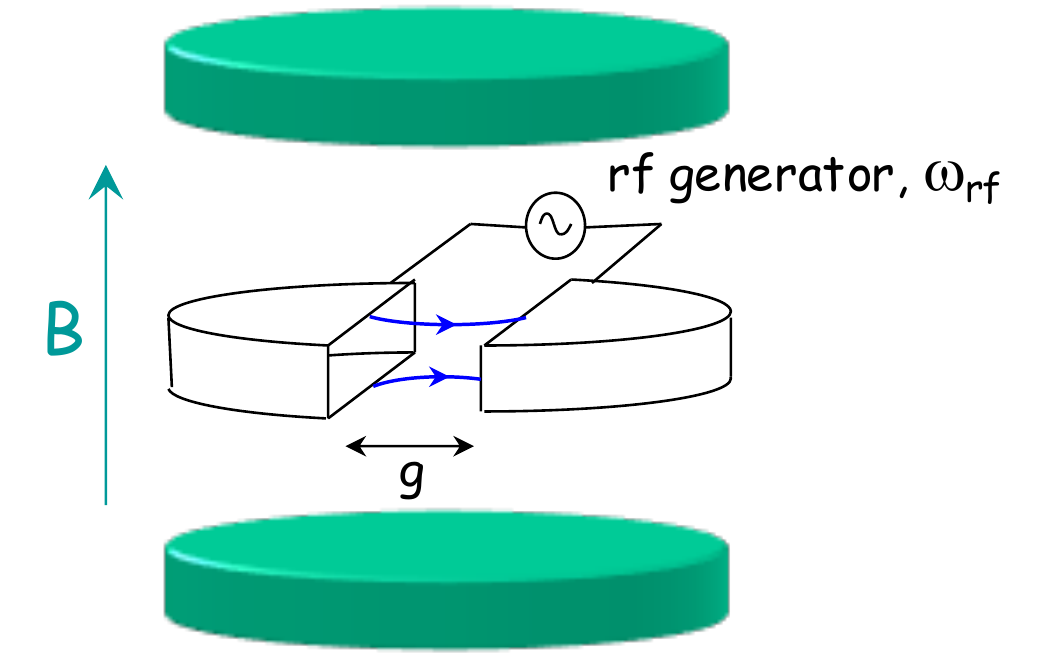}\hskip1cm
\includegraphics[width=0.45\textwidth]{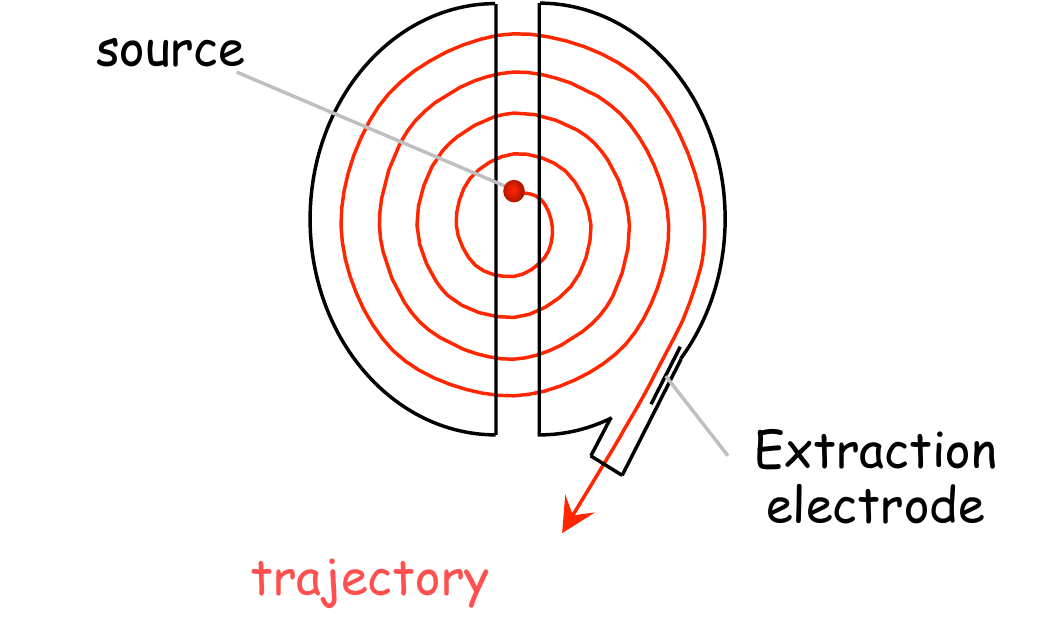}}
 \caption{\label{fig:cyclotron} Schematic view of a cyclotron and the particle trajectory in a top view (on the right)}
\end{figure}

Another type of circular accelerator based on the principle of time varying fields is the {\em cyclotron}.
The cyclotron has two hollow `D'-shape electrodes in a constant magnetic field $B$ (see Fig.~\ref{fig:cyclotron}).
When a particle is generated at the source in the centre, it is accelerated by the electric field between the electrodes. It enters an electrode and, while it is shielded from the electric field, the polarity of the field in the gap is reversed. If the phase of the rf is correct, the particle is accelerated again in the gap and enters the other electrode. The magnetic field creates a spiralling trajectory of the particle. As the particle becomes faster, the orbit radius gets bigger but the revolution frequency does not depend on the radius, as long as the particle is not relativistic.

So, the synchronism condition is that the rf period has to correspond to the revolution period:
\begin{equation}
T_{\rm rf} = 2\pi \rho / v_{\rm s}
\end{equation}
with the cyclotron frequency given by
\begin{equation}
\omega = \frac{q B}{m_0 \gamma}.
\end{equation}
As long as $v\ll c$ and $\gamma \approx 1$ the synchronism condition stays fulfilled. For higher energies, the particle will get out of phase with respect to the rf, even though there is still a range for the initial particle phase where a stable acceleration is possible.

In order to keep synchronism at higher energies, one has to decrease the radio frequency during the acceleration cycle according to the relativistic $\gamma(t)$ of the particle as
\begin{equation}
\omega_{\rm rf}(t) = \omega(t) = \frac{q B}{m_0 \gamma(t)},
\end{equation}
which leads to the concept of a {\em synchrocyclotron}, which can accelerate protons up to around 500~MeV. Here a new limitation occurs due to the size of the magnet. 
More details can be found in~\cite{bib:CAS94-LeDuff}.

\section{Synchrotron}
\begin{figure}[!b]
 \centering\includegraphics[width=0.55\textwidth]{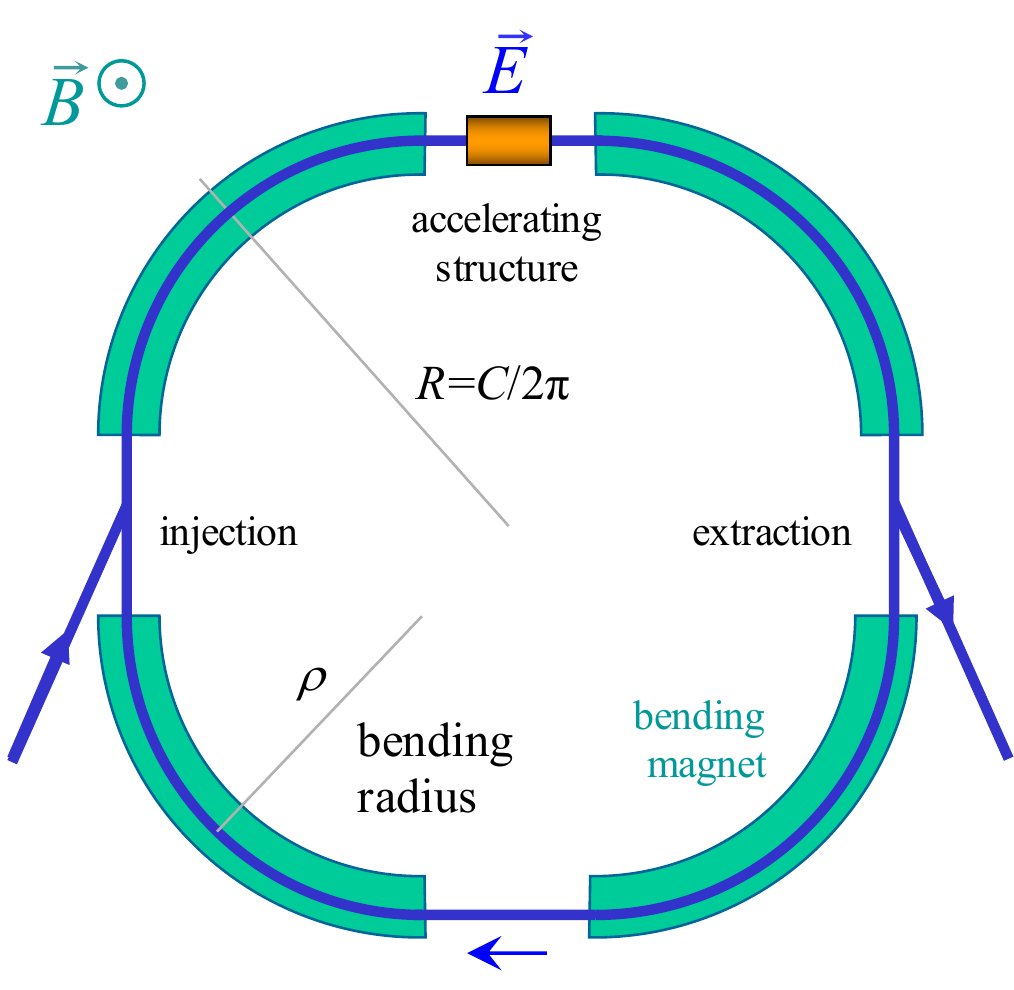}
 \caption{\label{fig:synchrotron} Schematic view of a synchrotron}
\end{figure}

A {\em synchrotron} (see Fig. \ref{fig:synchrotron}) is a circular accelerator in which the nominal particle trajectory is maintained at a
constant physical radius by varying both the magnetic field and the rf  to follow the energy variation.
In this way, the aperture of the vacuum chamber and the magnets can be kept small.

The rf needs to be synchronous to the revolution frequency. To achieve synchronism, the synchron\-ous particle needs to arrive at the cavity again after one turn with the same phase. This implies that the angular rf frequency, $\omega_{\rm rf} = 2\pi f_{\rm rf}$, must be an integer multiple of the angular revolution frequency~$\omega$:
\begin{equation}
\omega_{\rm rf} = h \, \omega \, ,
\end{equation}
where $h$ is an integer called the {\it harmonic number}. As a consequence, the number of stable synchron\-ous particle locations equals the harmonic number $h$. These locations are equidistantly spaced around the circumference of the accelerator. All synchronous particles will have the same nominal energy and will follow the nominal trajectory.

During the energy ramping, the magnetic field has to change in order to keep the radius constant, while the rf frequency needs to follow the change of the revolution frequency.
The time derivative of the momentum,
\begin{equation}
	p = q B \rho \, ,
\end{equation}
yields (when keeping the bending radius $\rho$ constant)
\begin{equation}
\der{p}{t} = q \rho \dot{B} \, .
\end{equation}
For one turn in the synchrotron, this results in
\begin{equation}
(\Delta p)_\mathrm{turn} = q \rho \dot{B} T = \frac{2 \pi q \rho R \dot{B}}{v} \: ,
\end{equation}
where the physical radius $R$ of the machine is defined by the circumference $C$ as $R = C / (2\pi)$.

From Eq.~\ref{eq:de} follows that
$\Delta E = v \Delta p$, and this energy gain per turn for the synchronous particle is provided by the rf system, 
so\footnote{In case of other effects changing the energy (synchrotron radiation, impedance, etc.), these would have to be added in the equation.} 
\begin{equation}
(\Delta E)_\mathrm{turn} = 2 \pi q \rho R \dot{B} = (\Delta E)_{\rm s} = q \hat{V} \sin \phi_{\rm s} \, .
\end{equation}

From this relation it can be immediately seen that the stable phase for the synchronous particle, the {\em synchronous phase $\phi_{\rm s}$}, changes during the acceleration, when the magnetic field $B$ changes, as
\begin{equation}
\sin \phi_{\rm s} = 2 \pi \rho R \, \frac{\dot{B}}{\hat{V}_{\rm rf}} \qquad \mathrm{or} \qquad \phi_{\rm s} = \arcsin\left( 2 \pi \rho R \, \frac{\dot{B}}{\hat{V}_{\rm rf}} \right) \, .
\end{equation}

As mentioned previously, the rf has to follow the change of revolution frequency and will increase during acceleration as
\begin{equation}
\frac{f_{\rm rf}}{h} = f = \frac{v(t)}{2\pi R} = \frac{1}{2\pi}\, \frac{qc^2}{E_0(t)} \,\frac{\rho}{R}\, B(t) \, .
\end{equation}

Since $E^2 = E_0^2 + p^2 c^2$,
the rf must follow the variation of the $B$ field with the relation
\begin{equation}
\frac{f_{\rm rf}}{h}  = \frac{c}{2\pi R} \sqrt{ \frac{B(t)^2}{[m_0c^2 / (q c \rho)]^2 + B(t)^2} } \, .
\end{equation}
So the rf frequency evolution can be directly calculated from the magnetic cycle.   
Obviously, this asymptotically approaches $f \rightarrow {c}/({2\pi R})$ for highly relativistic particles, as  $v \rightarrow c$\, (the field $B$ becomes large compared to $m_0c^2 / (q c \rho)$ ).

\subsection{Dispersion effects in a synchrotron}
If a particle is slightly shifted in momentum, it will have a different velocity and also a different orbit and orbit length.
We can define two parameters: 
\begin{itemize}
\item the {\it momentum compaction factor} $\alpha_\mathrm{c}$, which is the relative change in orbit length $C$ with momentum, given by 
\begin{equation}
\alpha_\mathrm{c} = \frac{\Delta C/C}{\Delta p / p} \, ;
\end{equation}
\item the {\it slip factor} $\eta$, which is the relative change in revolution period with momentum, given by
\begin{equation}
\eta = \frac{\Delta T/T}{\Delta p / p} \, . 
\end{equation}
\end{itemize}
(Remark: This is sometimes defined in the literature with the opposite sign.\footnote{I used the opposite sign convention myself in the past. I changed it to the present one to be consistent with other popular text books and simulation codes.})

\begin{figure}[!b]
\begin{tabular}{p{0.4\textwidth} p{0.4\textwidth}}
  \vspace{0pt}\hspace{1cm} \includegraphics[width=0.3\textwidth]{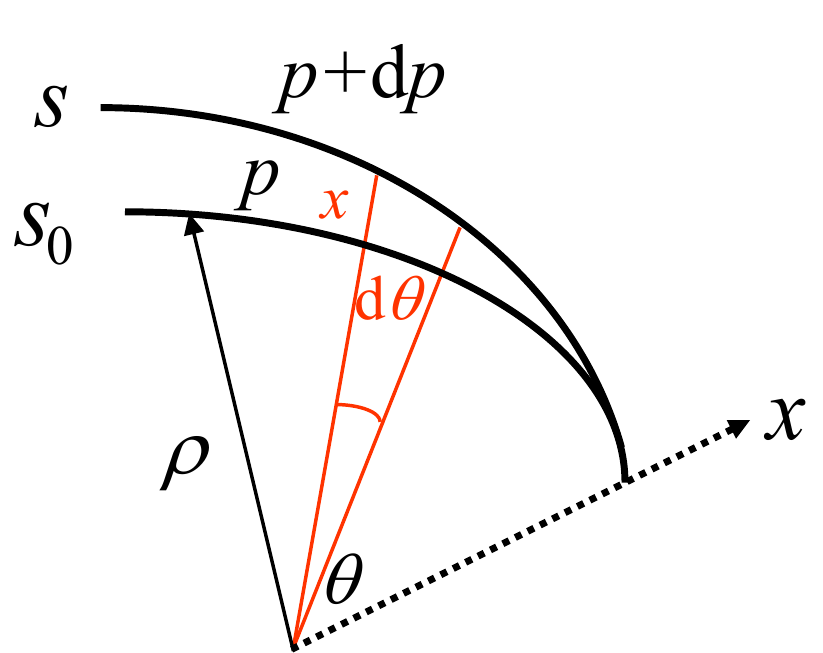} &
 \vspace{20pt}\begin{eqnarray}
\mathrm{d}s_0 & = & \rho \, \mathrm{d}\theta \, , \nonumber\\
\mathrm{d}s & = & ( \rho + x ) \, \mathrm{d}\theta \, . \nonumber
\end{eqnarray}
\end{tabular}
 \caption{\label{fig:orbit} Orbit length change}
\end{figure}

Let us consider the change in orbit length (see Fig. \ref{fig:orbit}). The relative elementary path-length difference $\mathrm{d}l$ for a particle with momentum $p + \mathrm{d}p$ is
\begin{equation}
\der{l}{s_0} = \der{s - \mathrm{d}s_0}{s_0} = \frac{x}{\rho} = \frac{D_x}{\rho} \frac{\mathrm{d}p}{p} \, ,
\end{equation}
where $D_x = \mathrm{d}x / (\mathrm{d}p/p)$ is the {\it dispersion function} from the transverse beam optics.

This leads to a total change $\mathrm{d}C$ in the circumference $C$ of
\begin{equation}
\mathrm{d}C = \int_C \mathrm{d}l = \int \frac{x}{\rho} \,\mathrm{d}s_0 = \int \frac{D_x}{\rho} \frac{\mathrm{d}p}{p} \,\mathrm{d}s_0 \, ,
\end{equation}
so that
\begin{equation}
\alpha_\mathrm{c} = \frac{1}{C} \int \frac{D_x}{\rho} \,\mathrm{d}s_0 \,
.
\end{equation}

Since $\rho = \infty$ in the straight sections, we get
\begin{equation}
\alpha_\mathrm{c} = \frac{\langle D_x \rangle_{\rm m}}{R} \, ,
\end{equation}
where the average $\langle \:\cdot\: \rangle_{\rm m}$ is considered over the bending magnets only.

Given that the revolution period is $T =  C/(\beta c)$, the relative change is (using the definition of the momentum compaction factor and \footnote{\begin{equation}
p = mv = \beta \gamma \frac{E_0}{c} \;\Rightarrow \; \frac{\rd p}{p} =  \frac{\rd\beta}{\beta} + \frac{\rd(1-\beta^2)^{-1/2}}{(1-\beta^2)^{-1/2}} = \underbrace{\left( 1-\beta^2 \right)^{-1}}_{\displaystyle\gamma^2} \frac{\rd\beta}{\beta} \, .
\end{equation}})
\begin{equation}
\frac{\rd T}{T} = \frac{\rd C}{C} - \frac{\rd\beta}{\beta} =  \alpha_\mathrm{c} \frac{\rd p}{p} - \frac{\rd\beta}{\beta} 
= \left(\alpha_\mathrm{c} - \frac{1}{\gamma^2} \right) \frac{\rd p}{p} \, ,
\end{equation}
which means that the slip factor $\eta$ is given by

\begin{equation}
\eta =\alpha_\mathrm{c} - \frac{1}{\gamma^2} \, .
\end{equation}

Obviously, there is one energy with a given $\gamma_{\rm t}$ for which $\eta$ becomes zero, meaning that there is no change in the revolution frequency for particles with a small momentum deviation. This so-called {\it transition energy} is a property of the transverse lattice, with
\begin{equation}
\gamma_{\rm t} = \frac{1}{\sqrt{\alpha_\mathrm{c}}} \: .
\end{equation}

From the definition of $\eta$, it is clear that an increase in momentum gives
the following.\begin{itemize}
\item {\it Below transition} energy $(\eta < 0)$: a higher revolution frequency. The increase in the velocity of the particle is the dominating effect.
\item {\it Above transition} energy $(\eta > 0)$: a lower revolution frequency. The particle has a velocity close to the speed of light; this velocity does not change significantly any more. Thus, here the effect of the longer path-length dominates (for the most common case of transverse lattices with a positive momentum compaction factor,  $\alpha_\mathrm{c}>0$).
\end{itemize}

At transition, the velocity change and the path-length change with momentum compensate for each other,
so the revolution frequency there is independent of the momentum deviation.
As a consequence, the longitudinal motion stops and the particles in the bunch will not change their phase.
Particles that are not at the synchronous phase will get the same non-nominal energy gain in each turn and will accumulate an energy error that will increase the longitudinal emittance and can lead to a loss of the particle due to dispersive effects. Therefore, transition has to be passed quickly to minimize the emittance increase and the losses.

Electron synchrotrons do not need to cross transition. Owing to the relatively small rest mass of the electron, the relativistic gamma factor is so large
that  the injection energy is already greater than the transition energy. Hence, the electrons will stay above transition during the whole acceleration cycle.

\begin{figure}[!b]
 \centering\includegraphics[width=0.9\textwidth]{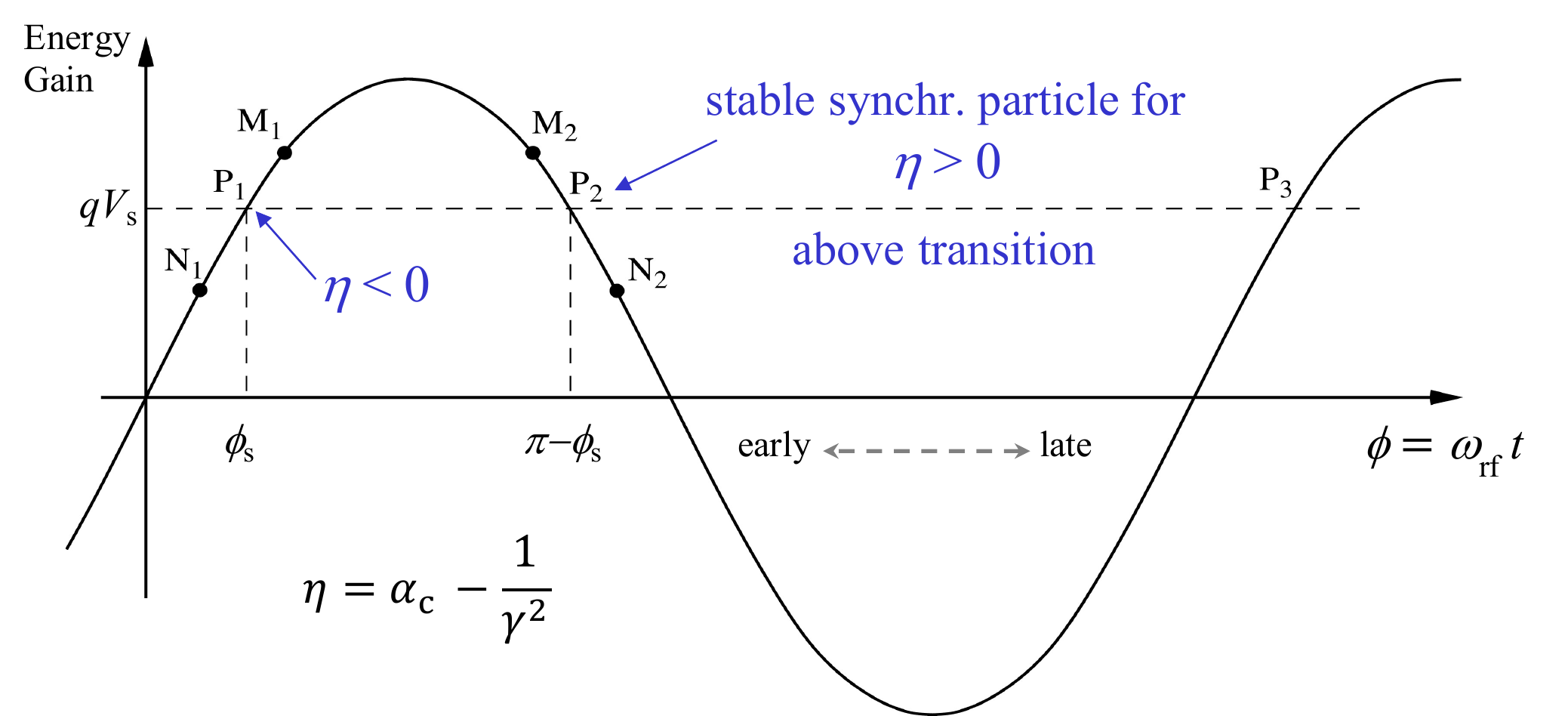}
 \caption{\label{fig:phase-stability} Energy gain as a function of particle phase; oscillations are stable around the synchronous-phase particle P$_1$ below transition and around the synchronous-phase particle P$_2$ above transition.}
\end{figure}

Since the changes in revolution frequency with momentum are opposite below and above transition, this completely alters the range for stable oscillations (see Fig.~\ref{fig:phase-stability}).
Let us consider a particle passing through an accelerating structure at a certain rf phase $\phi_{\rm s}$, where by design the energy gain is such that the particle reaches the structure again after one turn with the same phase $\phi_{\rm s}$.
This is illustrated in Fig.~\ref{fig:phase-stability} by the points P$_1$, P$_2$, and P$_3$.

Below transition, a particle N$_1$ which arrives earlier compared to P$_1$ will gain less energy and
its revolution time will become larger. At the
next turn, it will appear closer in time to particle P$_1$. The effect is opposite for particle M$_1$, which will gain
more energy and reduce its delay compared to P$_1$. So, the points P$_1$, P$_3$, etc are stable points
for the longitudinal oscillation below transition since particles slightly away from them will experience forces that will
reduce their deviation. On the contrary, it can be seen that the point P$_2$ is an unstable point
in the sense that particles slightly away from this point will deviate even more in the next turns.

So, as in the linac case, the oscillation is stable for a particle that is on the rising slope of the rf field when we are below transition. Above transition, the oscillation is unstable on the rising slope, and the stable region for oscillations is on the falling slope. A particle that arrives too early (M$_2$) will gain more energy, and the revolution time will increase owing to the predominant effect of the longer path; thus, the particle will arrive later on the next turn, closer to the synchronous phase. Similarly, a particle that arrives late (N$_2$) will gain less energy and travel a shorter orbit, also moving towards the synchronous phase.

\begin{figure}[!bt]
 \centering\includegraphics[width=0.6\textwidth]{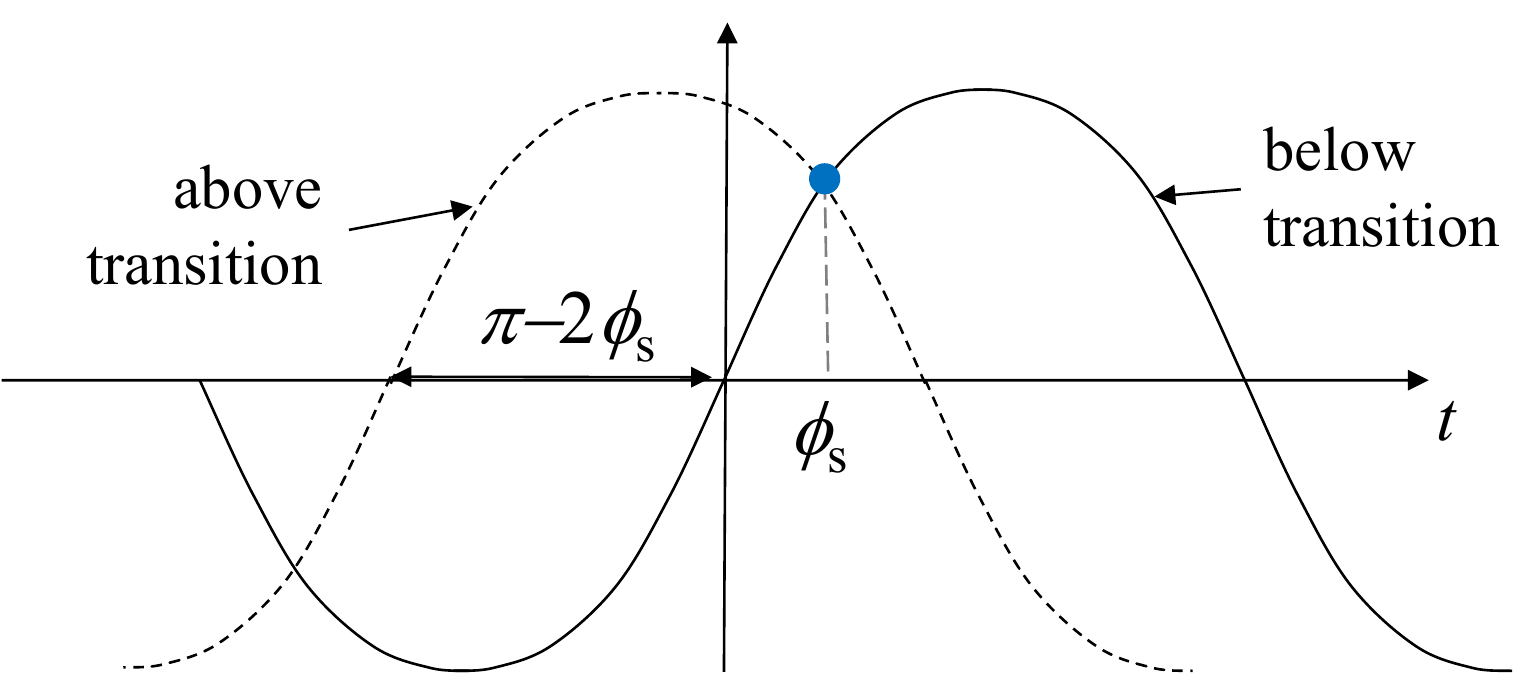}
 \caption{\label{fig:transition} The synchronous phase changes from a rising to a falling slope as the energy crosses transition.}
\end{figure}

Crossing transition during acceleration makes the previous stable synchronous phase unstable. Below transition, it is stable on the rising slope of the rf; above transition, the synchronous phase is on the falling slope. Consequently, the rf system needs to make a rapid change of rf phase when crossing the transition energy---a `phase jump', as indicated in Fig.~\ref{fig:transition}---otherwise the particles in the bunch become dispersed, have a wrong energy gain, and eventually get lost. 
The rf phase needs to change by $\pi - 2\phi_{\rm s}$. 

A method to improve transition crossing is to change the transverse optics when the energy almost reaches $\gamma_{\rm t}$ for an optics with a larger momentum compaction factor $\alpha_\mathrm{c}$. Hence, $\gamma_{\rm t}$ is decreasing at the same time as the energy is increasing, and so the time with an energy close to transition can be reduced.

\subsection{Equations of longitudinal motion in a synchrotron}
\begin{figure}[!b]
\begin{center}
\begin{minipage}[t]{0.3\textwidth}
\vspace*{10pt}\includegraphics[width=0.7\textwidth]{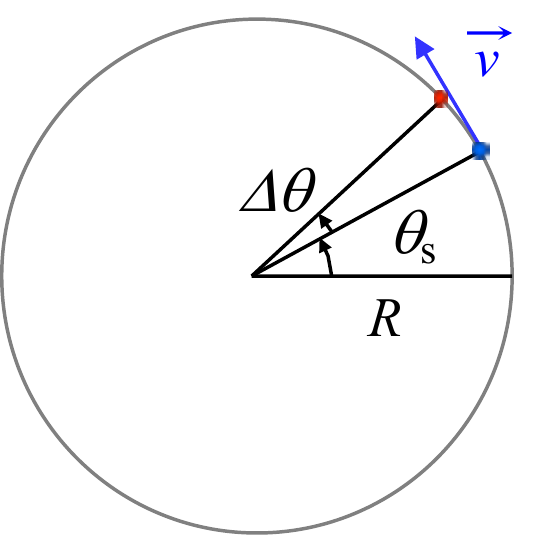}
\end{minipage}\hspace*{-5mm}
\begin{minipage}[t]{0.49\textwidth}
\vspace*{30pt}\begin{tabular}{ll}
particle rf phase & $\Delta\phi = \phi - \phi_{\rm s}\:$, \\
particle momentum & $\Delta p = p - p_0\:$, \\
particle energy & $\Delta E = E - E_0\:$, \\
azimuth angle& $\Delta\theta = \theta - \theta_{\rm s}\: .$
\end{tabular}
\end{minipage}
\end{center}
\caption{Variables with respect to synchronous particle}
\label{fig:Fred}
\end{figure}

As we have identified the special role of the synchronous particle, we want to look at the oscillations with respect to it, and we express the variables with respect to the synchronous particle (denoted by the subscripts `s' or `0'), as shown in Fig.~\ref{fig:Fred}.

Since the rf is a multiple of the revolution frequency ($f_\mathit{\rm rf} = h f$), the rf phase change $\Delta\phi$ 
is
\begin{equation}
\Delta\phi = -h \, \Delta\theta \quad \text{with }\: \theta = \int \omega\, \rd t \, .
\end{equation}
The minus sign for the rf phase arises from the fact that a particle that is ahead arrives earlier, i.e.\ at a smaller rf phase.

For a given particle with respect to the reference particle, the change in angular revolution frequency is
\begin{equation}
\Delta\omega = \der{}{t} \left( \Delta\theta \right) = - \frac{1}{h} \der{}{t} \left( \Delta\phi \right) = - \frac{1}{h} \der{\phi}{t} \: .
\end{equation}

Since (with $\omega_0$ being the revolution frequency for the synchronous particle)
 \[
 \displaystyle\eta = - \frac{p_0}{\omega_0} \left( {\frac{\mathrm{d}\omega}{\mathrm{dp}}} \right)_{\!{\rm s}},
\quad E^2 = E_0^2 + p^2 c^2 ,\quad  \mathrm{and} \quad \Delta E = v_0 \Delta p = \omega_0 R \Delta p \, , 
 \]
one gets the first-order equation
\begin{equation}
\frac{\Delta E}{\omega_0} =  \frac{p_0 R}{h \eta \omega_0} \der{(\Delta\phi)}{t} =  \frac{p_0 R}{h \eta \omega_0} \, \dot{\phi} \, . \label{eq:first1}
\end{equation}
The second first-order equation follows from the energy gain of a particle:
\begin{equation}
\der{E}{t} = \frac{\omega}{2\pi} \,q \hat{V} \sin\phi \, ,
\end{equation}
so that the time derivative of the energy difference becomes (in first order approximation)
\begin{equation}
\der{}{t} \left( \frac{\Delta E}{\omega_0} \right) = \frac{q \hat{V}}{2\pi} \bigl( \sin\phi - \sin\phi_{\rm s} \bigr)\: . \label{eq:first2}
\end{equation}

Combining the two first-order equations (Eqs. (\ref{eq:first1}) and (\ref{eq:first2})) leads to
\begin{equation}
\der{}{t} \left[ \frac{-R p_0}{h \eta \omega_0} \der{\phi}{t} \right] + \frac{q \hat{V}}{2\pi}  ( \sin\phi - \sin\phi_{\rm s}) = 0 \, .
\label{eq:gen2}
\end{equation}
This second-order equation is non-linear, and the parameters within the bracket are, in general, slowly varying in time.

When we assume constant parameters $R, p_0, \omega_0$, and $\eta$, we get
\begin{equation}
\ddot\phi + \frac{\Omega_{\rm s}^2}{\cos\phi_{\rm s}}  \left( \sin\phi - \sin\phi_{\rm s} \right) = 0 \qquad \text{with } \: 
\Omega_{\rm s}^2 = \frac{-h \eta \omega_0 q \hat{V} \cos\phi_{\rm s}}{2\pi R p_0}
=  \omega_0^2 \frac{-h \eta q \hat{V} \cos\phi_{\rm s}}{2\pi \beta^2 E_0} \label{eq:osc}
\end{equation}
and, for small phase deviations from the synchronous particle,
\begin{equation}
\sin\phi - \sin\phi_{\rm s} = \sin(\phi_{\rm s}+\Delta\phi) - \sin\phi_{\rm s} \approxeq \cos\phi_{\rm s} \Delta\phi \, ,
\end{equation}
so that we end up with the equation of a harmonic oscillator:
\begin{equation}
\ddot\phi + \Omega_{\rm s}^2 \Delta\phi = 0 \, ,
\end{equation}
where $\Omega_{\rm s}$ is the synchrotron angular frequency.

\begin{figure}[!htb]
 \centering\includegraphics[width=0.85\textwidth]{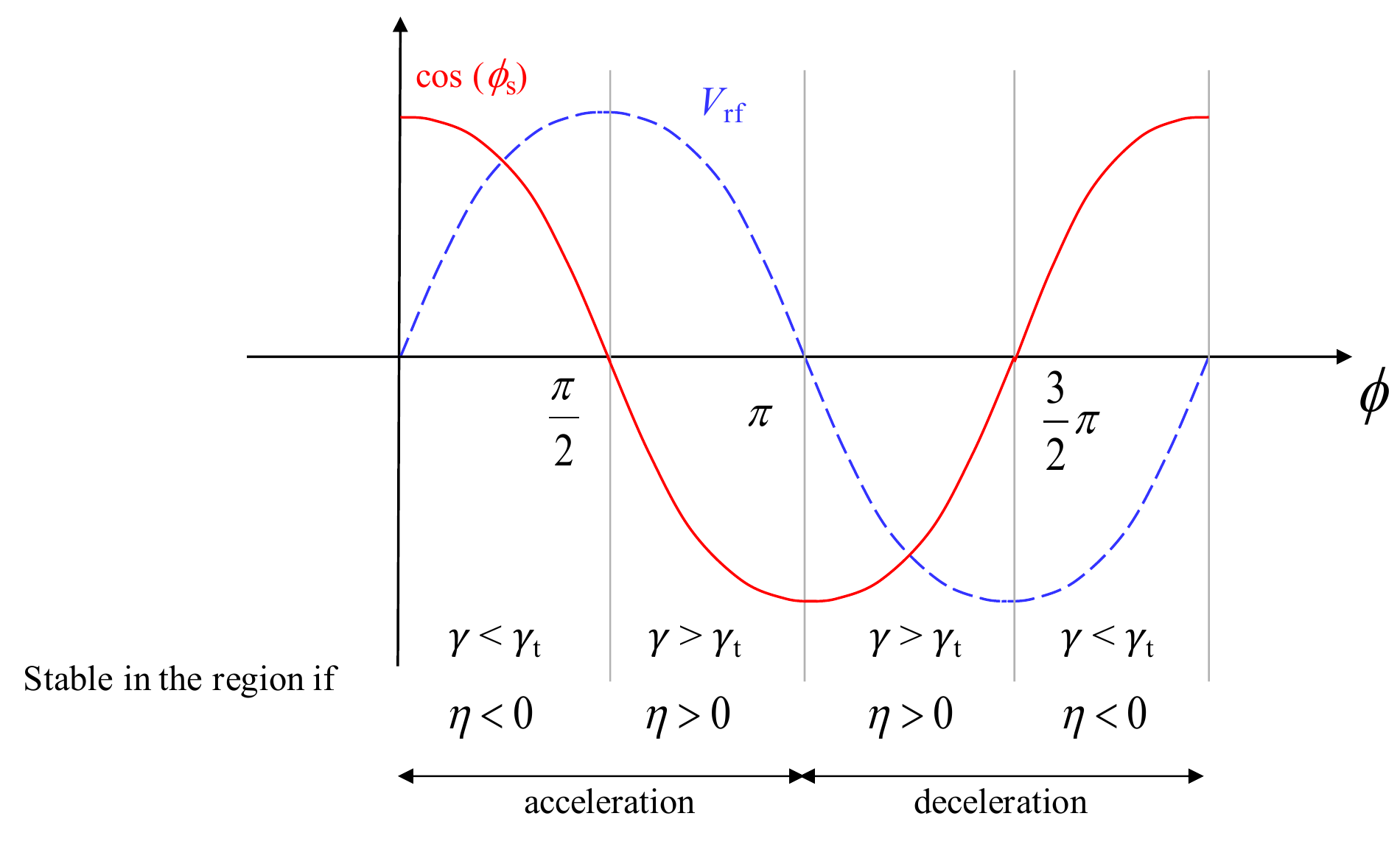}
 \caption{\label{fig:stability-regions} Stability regions as a function of particle phase, depending on the energy with respect to transition
}
\end{figure}

Stability is obtained when $\Omega_{\rm s}$ is real so that $\Omega_{\rm s}^2$ is positive. Since most terms in the expression for $\Omega_{\rm s}^2$ are positive (and choosing $\hat{V}$ such that $q\hat{V}>0$), this condition reduces to
\begin{equation}
\eta \cos\phi_{\rm s} < 0\: ,
\end{equation}
and the stable region for the synchronous phase depends on the energy relative to the transition energy, as we have seen from our previous argument. The conditions for stability are summarized in Fig.~\ref{fig:stability-regions}.

The \emph{synchrotron tune} $Q_{\rm s}$ is defined as
\begin{equation}
Q_{\rm s} = \Omega_{\rm s} / \omega_0 
\end{equation}
and corresponds to the number of synchrotron oscillations per revolution in the synchrotron. Generally $Q_{\rm s} \ll 1$, as it  typically
takes of the order of several tens to several hundreds of turns to complete one synchrotron oscillation.

For larger phase (or energy) deviations from the synchronous particle, we can multiply Eq. (\ref{eq:osc}) by $\dot\phi$ and integrate it, obtaining an invariant of motion
\begin{equation}
\frac{\dot\phi^2}{2} - \frac{\Omega_{\rm s}^2}{\cos\phi_{\rm s}}  \left( \cos\phi + \phi \sin\phi_{\rm s} \right) = I \, ,
\label{eq:invariant}
\end{equation}
which for small amplitudes of $\Delta\phi$ reduces to
\begin{equation}
\frac{\dot\phi^2}{2} + \Omega_{\rm s}^2 \,\frac{(\Delta\phi)^2}{2} = I' \,
.
\end{equation}
Similar equations can be derived for the second variable $\Delta E \propto \mathrm{d}\phi
/ \mathrm{d}{t}$.

\begin{figure}[!htb]
 \centering\includegraphics[width=0.6\textwidth]{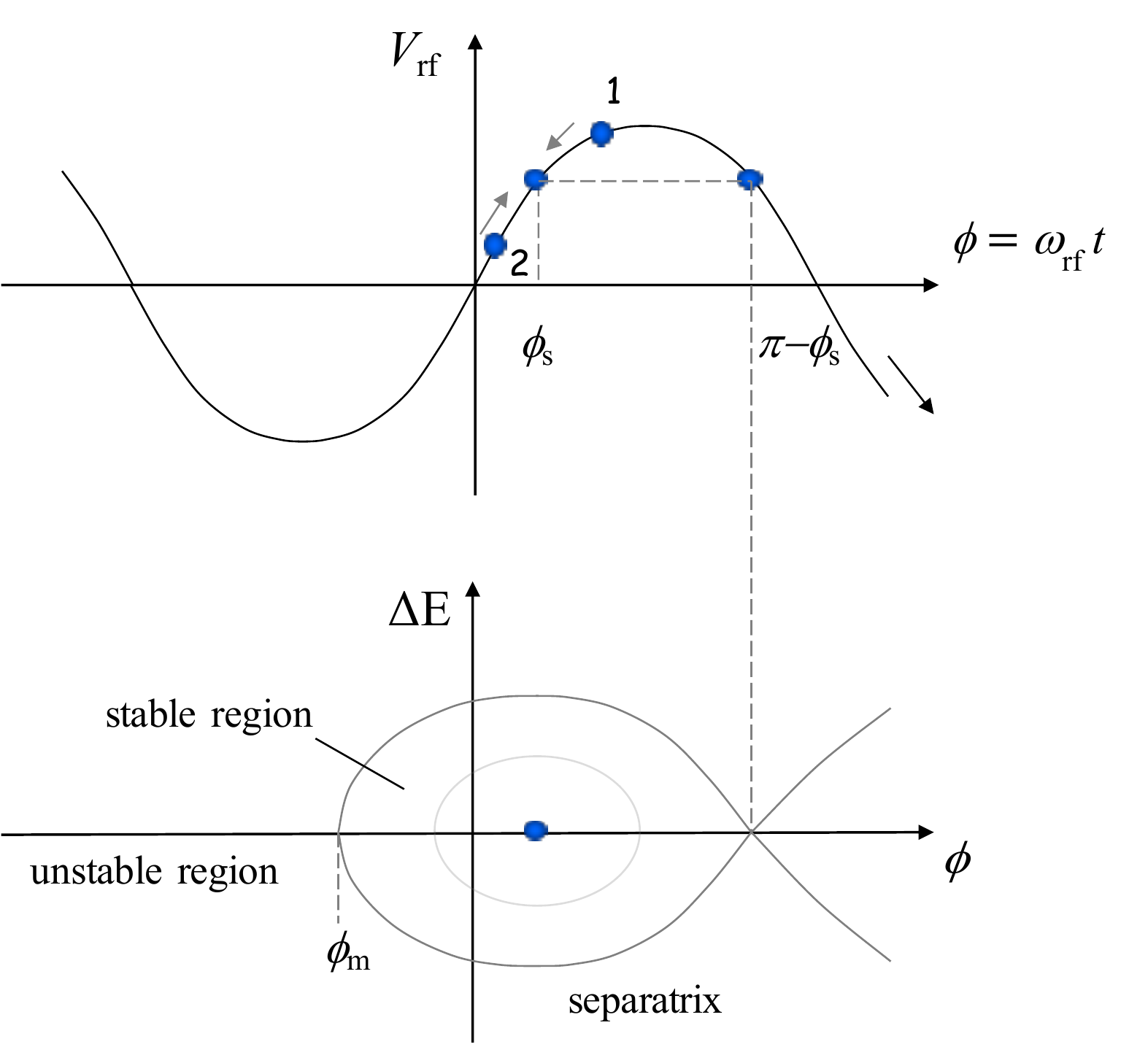}
 \caption{\label{fig:acc-bucket} Top graph shows rf voltage ($=$ energy gain) as a function of particle phase; bottom diagram is the corresponding phase-space picture for the case with a synchronous phase of $\phi_{\rm s}$}
\end{figure}

In order to visualize the the energy--phase oscillations, one typically uses a {\em phase-space} plot. 
The horizontal axis is phase (or time, depending on the variables used), the vertical axis is the energy deviation (or momentum deviation, etc.).
Figure~\ref{fig:acc-bucket} shows an example of a phase-space plot.

When we look at the energy gain with respect to the synchronous particle at $\phi_{\rm s}$ in Fig.~\ref{fig:acc-bucket},
we see that the restoring force goes to zero when $\phi$ reaches $\pi - \phi_{\rm s}$, and it becomes non-restoring beyond (both below and above transition).
Hence $\pi - \phi_{\rm s}$ is an extreme amplitude for a stable motion, which has a closed trajectory in phase space. 
This phase-space trajectory separates the region of stable motion from the unstable region; it is called the {\em separatrix}. 
The area within this separatrix is called the {\em rf bucket\/} and  corresponds to the maximum acceptance in phase space for a stable motion.
Particles located in phase-space within this separatrix will move around on closed phase-space trajectories.

Since we have found an invariant of motion in Eq.~\ref{eq:invariant}, we can write the equation for the separatrix by calculating it at the phase $\phi = \pi - \phi_{\rm s}$ where $\dot\phi = 0$:
\begin{equation}
\frac{\dot\phi^2}{2} - \frac{\Omega_{\rm s}^2}{\cos\phi_{\rm s}}  \left( \cos\phi + \phi \sin\phi_{\rm s} \right) = - \frac{\Omega_{\rm s}^2}{\cos\phi_{\rm s}}  \bigl( \cos(\pi -\phi_{\rm s}) + (\pi -\phi_{\rm s}) \sin\phi_{\rm s} \bigr) \: .
\label{eq:separatrix}
\end{equation}

From this we can calculate the second value, $\phi_{\rm m}$, where the separatrix crosses the horizontal axis, which is the other extreme phase for stable motion:
 \begin{equation}
 \cos\phi_{\rm m} + \phi_{\rm m} \sin\phi_{\rm s}  = \cos(\pi -\phi_{\rm s}) + (\pi -\phi_{\rm s}) \sin\phi_{\rm s} \, .
\end{equation}

It can be seen from the equation of motion that $\dot\phi$ reaches an extreme when $\ddot\phi = 0$, corresponding to $\phi = \phi_{\rm s}$. Putting this value into Eq. (\ref{eq:separatrix}) gives
\begin{equation}
\dot\phi^2_\mathrm{max} = 2 \,\Omega_{\rm s}^2 \bigl[ 2 + (2\phi_{\rm s}-\pi ) \tan\phi_{\rm s} \bigr] \: ,
\end{equation}
which translates into an acceptance in energy
\begin{equation}
\left( \frac{\Delta E}{E_0} \right)_\mathrm{max} = \pm \beta \sqrt{ \frac{-q \hat{V}}{\pi h \eta E_0} \,G(\phi_{\rm s}) } \; ,
\label{eq:energy_acc}
\end{equation}
where
\begin{equation}
G(\phi_{\rm s}) = 2 \cos\phi_{\rm s} + ( 2\phi_{\rm s}-\pi ) \sin\phi_{\rm s} \, .
\end{equation}

\begin{figure}[!hb]
 \centering\includegraphics[width=0.55\textwidth]{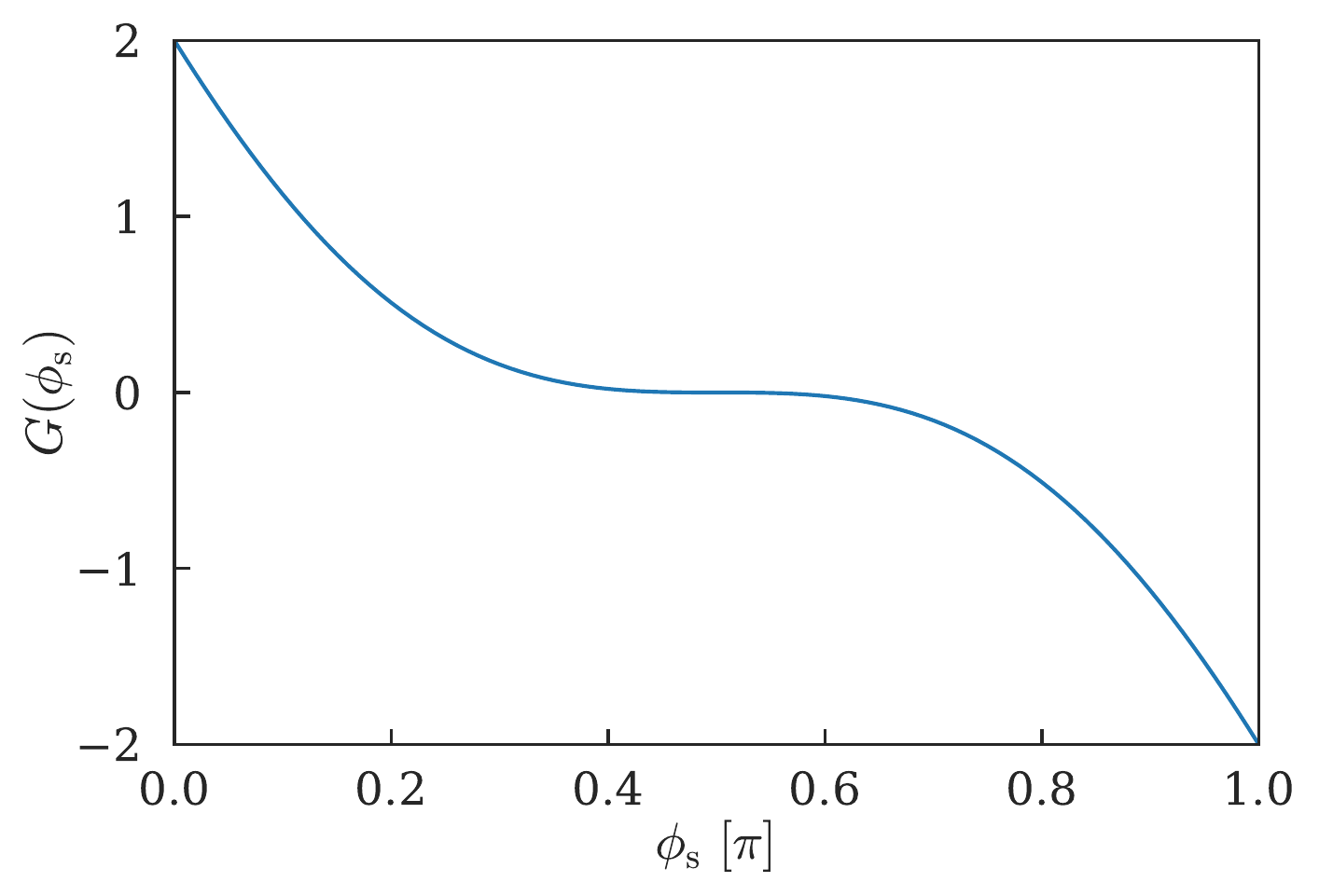}
 \caption{\label{fig:G} The function $G(\phi_{\rm s})$ for the energy acceptance in Eq.~(\ref{eq:energy_acc})}
\end{figure}

\begin{figure}[!htb]
 \centering\includegraphics[width=0.72\textwidth]{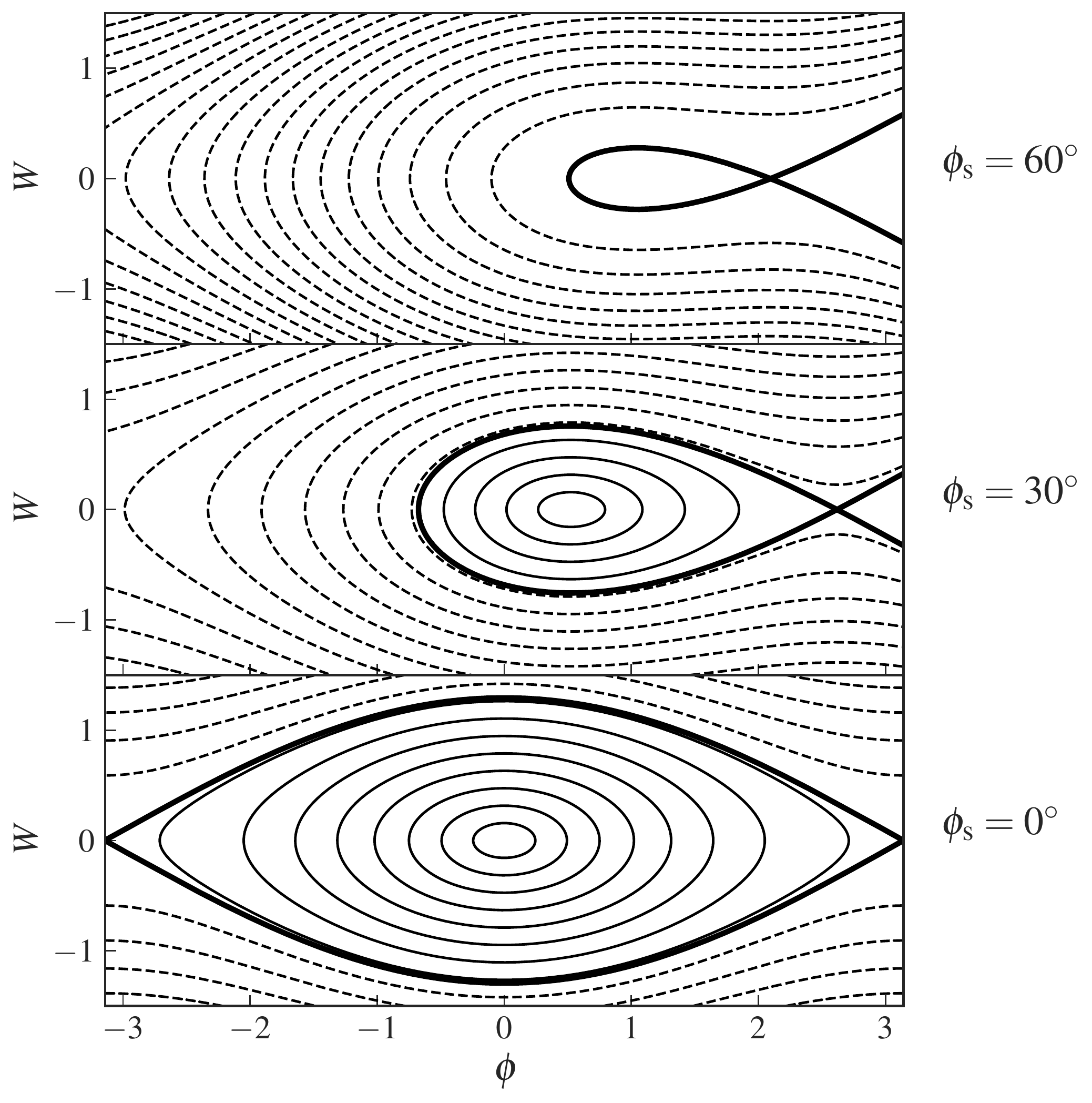}
 \caption{\label{fig:acceptance} Phase-space plots for different synchronous-phase angles $\phi_\mathrm{s}$ for otherwise identical parameters; in each plot, thin solid lines represent stable trajectories in phase space, dashed lines represent unstable trajectories, and the thick solid line is the separatrix.}
\end{figure}

This {\it rf acceptance} depends strongly on $\phi_{\rm s}$, as can be seen from the function $G(\phi_{\rm s})$ in Fig.~\ref{fig:G}, and plays an important role in the capture at injection and the stored beam lifetime.
The maximum energy acceptance in the bucket depends on the square root of the available rf voltage, $\hat{V}_{\rm rf}$.
The phase extension of the bucket is at a maximum for $\phi_{\rm s} = 0^\circ$ or $180^\circ$.
As the synchronous phase approaches $90^\circ$, the bucket size becomes smaller, as illustrated in Fig.~\ref{fig:acceptance}.

\subsection{Stationary bucket}
For a stationary bucket (constant energy, no acceleration), we have $\sin\phi_{\rm s} = 0$, so  $\phi_{\rm s} = 0$ or $\pi$. For the case of $\phi_{\rm s} = \pi$ (above transition, $\eta>0$) the equation of the separatrix simplifies to
\begin{equation}
\frac{\dot\phi^2}{2} + \Omega_{\rm s}^2 \cos\phi = \Omega_{\rm s}^2 \qquad \mathrm{or} \qquad \frac{\dot\phi^2}{2} = 2 \, \Omega_{\rm s}^2 \sin^2\frac{\phi}{2}
\; .
\end{equation}

At this point, it is convenient to introduce a new variable $W$ to replace the phase derivative $\dot\phi$:
\begin{equation}
W = \frac{\Delta E}{\omega_0} = \frac{p_0 R}{h \eta \omega_0} \,\dot\phi \, ,
\end{equation}
where $\omega_0$ is the revolution frequency of the synchronous particle. As we shall see later, this new variable is canonical. Different choices of canonical variables are possible and lead to slightly different equations, as, for example, in Ref.~\cite{bib:CAS2013-tecker}.

Introducing $\Omega_{\rm s}^2$ from Eq. (\ref{eq:osc}) leads to the following equation for the separatrix ($\eta>0$):
\begin{equation}
W = \pm \frac{R}{c} \sqrt{ \frac{2q \hat{V} E_0}{\pi h \eta}}\:  \sin\frac{\phi}{2} = \pm W_{\rm bk} \sin\frac{\phi}{2} \qquad \text{with } \;
W_{\rm bk} = \frac{R}{c} \sqrt{ \frac{2q \hat{V} E_0}{\pi h |\eta|}}
\; .
\end{equation}

Setting $\phi = \pi$ in the previous equation shows that $W_{\rm bk}$ is the maximum height of the bucket, which results in the maximum energy acceptance (with $\beta_0$ denoting the $\beta$ of the synchronous particle)
\begin{equation}
\Delta E_\mathrm{max} = \omega_0 W_{\rm bk} = \beta_0 \sqrt{ \frac{2q \hat{V}_{\rm rf} E_0}{\pi h |\eta|}} \; .
\end{equation}
Below transition ($\eta<0$) the expression for $W_{\rm bk}$ is identical\footnote{$\eta$ is negative but there is an additional minus sign under the square root. So $|\eta|$ is correct for both below and above transition.}. 

The bucket area $A_{\rm bk}$ is
\begin{equation}
A_{\rm bk} = 2 \int_0^{2\pi} W \,\mathrm{d}\phi \: .
\end{equation}
With $\int_0^{2\pi} \sin(\phi/2) \, \mathrm{d}\phi = 4$, one obtains
\begin{equation}
A_{\rm bk} = 8 W_{\rm bk} = 16 \frac{R}{c} \sqrt{ \frac{q \hat{V} E_0}{2 \pi h |\eta|}} 
= 16 \frac{\beta_0}{\omega_0} \sqrt{ \frac{q \hat{V} E_0}{2 \pi h |\eta|}} \; .
\end{equation}

As we have seen qualitatively before, the bucket area gets reduced for an accelerating bucket.
The reduction factor $\alpha(\phi_{\rm s})$ can be approximated by
\begin{equation}
\alpha(\phi_{\rm s}) \approx \frac{1-\sin\phi_{\rm s}}{1+\sin\phi_{\rm s}}
\end{equation}
\subsection{Bunch matching into the stationary bucket}
\begin{figure}[!b]
 \centering\includegraphics[width=0.47\textwidth]{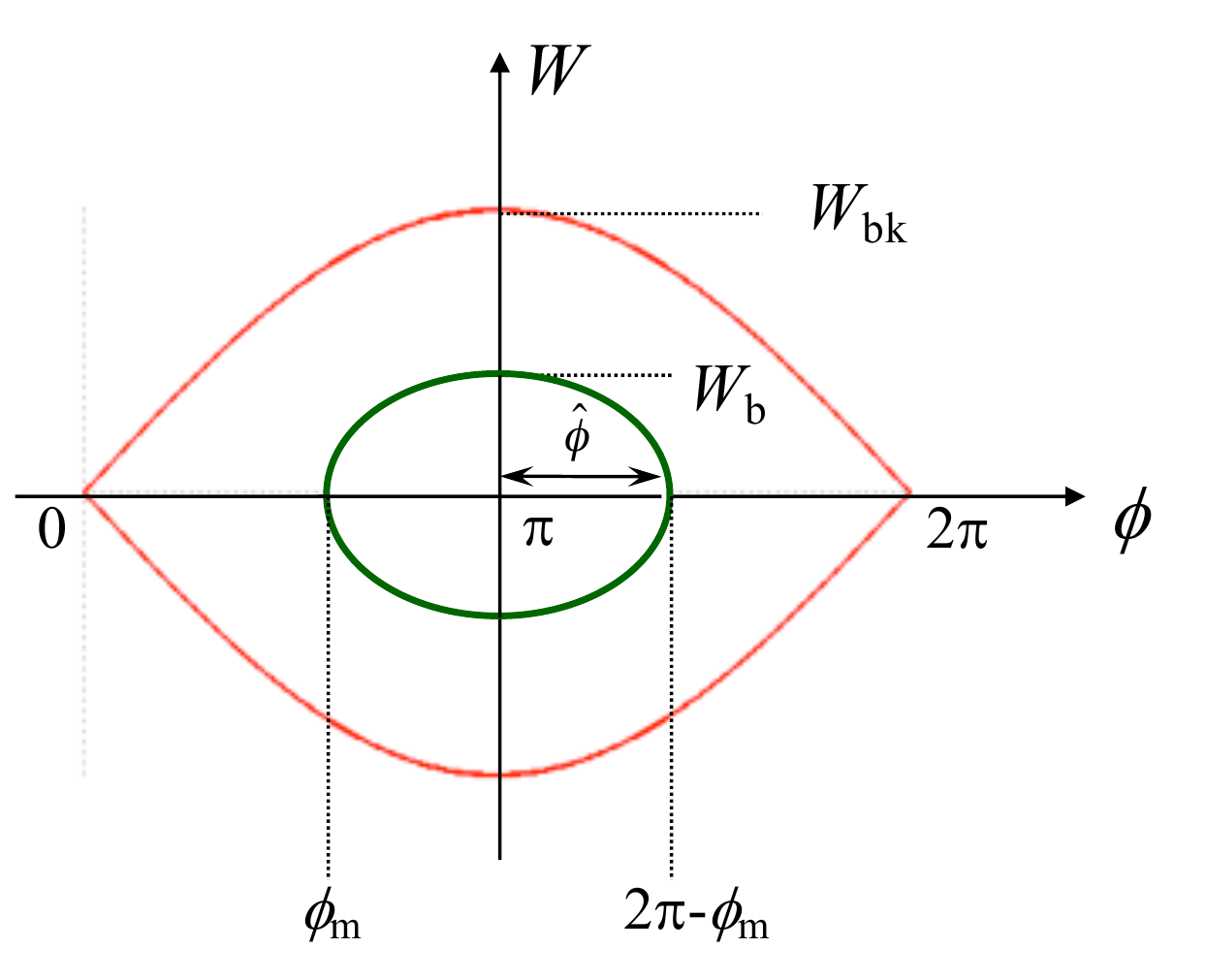}
 \caption{\label{fig:stationary-bucket} Phase-space plot for the separatrix of the stationary bucket and a trajectory inside}
\end{figure}

We can describe the motion of a particle inside the separatrix of the stationary bucket (here above transition)
by starting from the invariant of motion in Eq. (\ref{eq:invariant}) and setting $\phi_{\rm s} = \pi$:
\begin{equation}
\frac{\dot\phi^2}{2} + \Omega_{\rm s}^2 \,\cos\phi = I \, .
\end{equation}
The points $\phi_{\rm m}$ and $2\pi-\phi_{\rm m}$ where the trajectory crosses the horizontal axis are symmetrical with respect to $\phi_{\rm s} = \pi$ (see Fig. \ref{fig:stationary-bucket}). 
We can calculate the invariant for $\phi = \phi_{\rm m}$ and get
\begin{equation}
\frac{\dot\phi^2}{2} + \Omega_{\rm s}^2 \,\cos\phi = \Omega_{\rm s}^2 \,\cos\phi_{\rm m}
\qquad \Rightarrow \qquad
\dot\phi = \pm \Omega_{\rm s} \sqrt{2\, (\cos\phi_{\rm m} - \cos\phi)}\; ,
\end{equation}
\begin{equation}
W = \pm W_{\rm bk} \sqrt{\cos^2\frac{\phi_{\rm m}}{2} - \cos^2\frac{\phi}{2}} \qquad \Bigl( \mathrm{using } \; \cos\phi = 2 \cos^2\frac{\phi}{2} - 1 \Bigr)\; .
\end{equation}

Setting $\phi = \pi$ in the previous equation allows us to calculate the bunch height $W_{\rm b}$:
\begin{equation}
W_{\rm b} = W_{\rm bk} \,\cos\frac{\phi_{\rm m}}{2} = W_{\rm bk} \,\sin\frac{\hat\phi}{2}
\: ,
\end{equation}
with $\hat\phi = \pi - \phi_{\rm m}$ being the maximum phase amplitude for an oscillation around the synchronous phase. 

The corresponding maximum energy difference of a particle on this phase-space trajectory is
\begin{equation}
\left( \frac{\Delta E}{E_0} \right)_{\!{\rm b}} = \left( \frac{\Delta E}{E_0} \right)_{\rm rf} \cos\frac{\phi_{\rm m}}{2} = \left( \frac{\Delta E}{E_0} \right)_{\rm rf} \sin\frac{\hat\phi}{2} \; .
\end{equation}

When a particle bunch is injected into a synchrotron, the bunch has a given bunch length and energy spread. Each of the different particles will move along a phase-space trajectory that corresponds to its initial phase and energy. If the shape of the injected bunch in phase space matches the shape of a phase-space trajectory for the given rf parameters, the shape of the bunch in phase space will be maintained.

\begin{figure}[!b]
 \centering\includegraphics[width=0.9\textwidth]{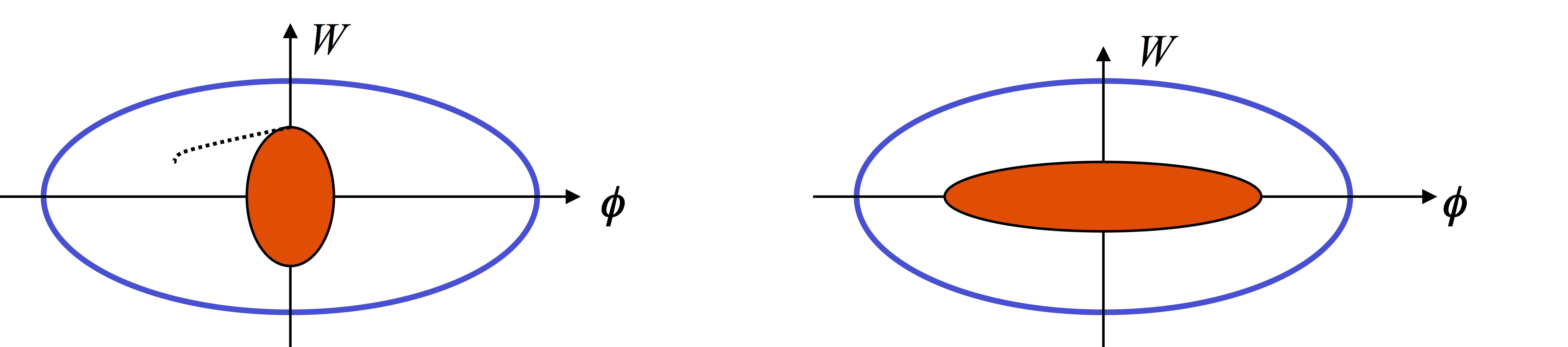}
 \caption{\label{fig:matching} Phase-space plots for a mismatched bunch one-quarter of a synchrotron period apart}
\end{figure}

If the shape does not match, it will vary during the synchrotron period. This is illustrated in Fig.~\ref{fig:matching} for a bunch that has a shorter bunch length and a larger energy spread compared with the phase-space trajectory. As the particles move along their individual trajectories, the bunch will be longer with a smaller energy spread after one-quarter of a synchrotron period, and it will regain the initial shape after one-half of a period. This effect can be used to manipulate the shape of the bunch in phase space and trade off bunch length against energy spread (so-called {\it bunch rotation}). When the rf voltage in matched conditions is suddenly increased, the bunch will be shorter after a quarter of a synchrotron period. In this way, it can be shortened for a transfer to a higher-frequency rf system.

\begin{figure}[!tb]
 \centering\includegraphics[width=0.85\textwidth]{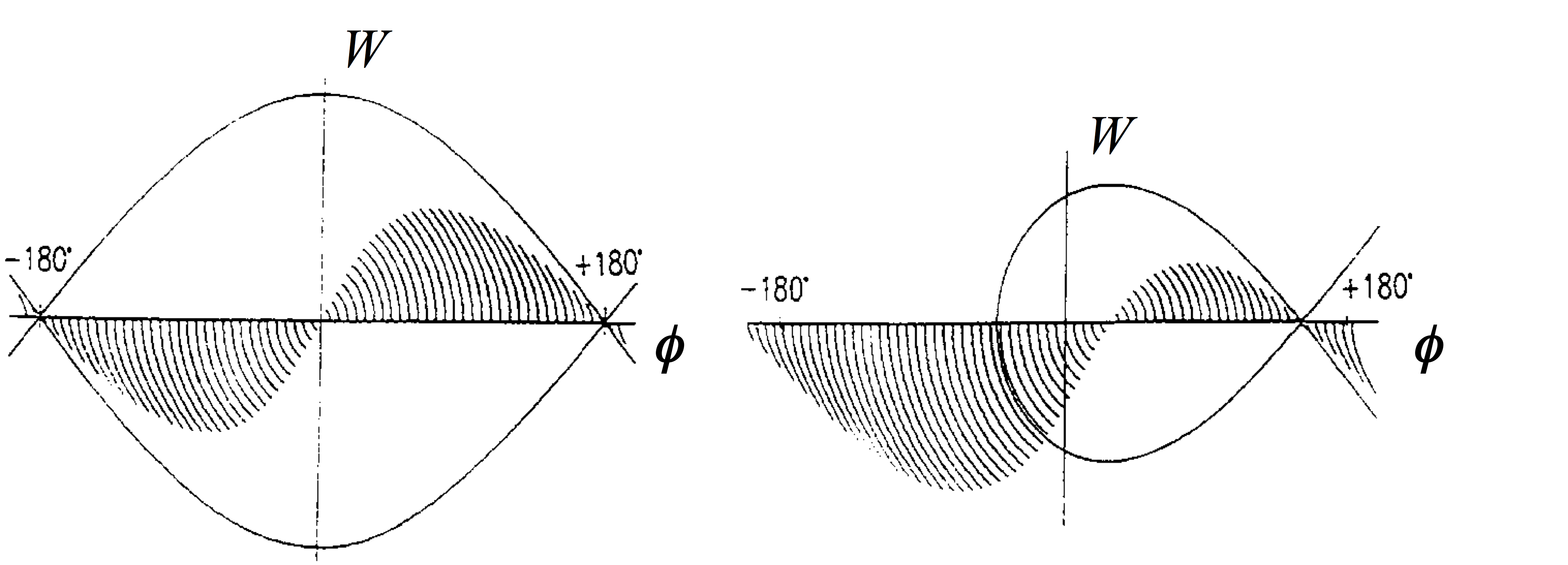}
 \caption{\label{fig:bucket-rotation} Phase-space trajectories for one-eighth of the synchrotron period: left, stationary bucket; right, accelerating bucket \cite{pirkl}.}
\end{figure}

Owing to the non-linear restoring force, the synchrotron period depends on the oscillation amplitude, and particles with larger amplitudes have a longer synchrotron period, as shown in Fig. \ref{fig:bucket-rotation}. This will eventually lead to {\it filamentation} of the bunch and an increase in  the longitudinal emittance.

The same phenomenon will occur when the bunch shape is matched to the bucket but there is an error in the phase or the energy. The different particles in the bunch
will perform their individual oscillations around the synchronous particle and will filament, leading to an increase in the longitudinal emittance. To avoid an emittance increase, it is important to match the phase, the energy, and the shape of the bunch and the bucket during the transfer.

\subsection{Potential energy function and Hamiltonian}
The longitudinal motion is produced by a force that can be derived from a scalar potential $U$:
\begin{align}
\der{^2\phi}{t^2} &= F(\phi) = - \frac{\partial U}{\partial\phi} \: ,
\\
U(\phi) &= - \int\limits_0^\phi F(\phi) \, \rd\phi = - \frac{\Omega^2_{\rm s}}{\cos\phi_{\rm s}} \,( \cos\phi + \phi\sin\phi_{\rm s} ) - F_0
\: .
\end{align}
The sum of the potential energy and the kinetic energy is constant and, by analogy, represents the total energy of a non-dissipative system:
\begin{equation}
\frac{\dot\phi^2}{2} + U(\phi) = F_0 \, .
\end{equation}

Since the total energy is conserved, we can describe the system as a Hamiltonian system. Different choices of the canonical variables are possible.
With the variable
\begin{equation}
W = \frac{\Delta E}{\omega_0} \, ,
\end{equation}
the two first-order equations of the longitudinal motion become
\begin{align}
\der{\phi}{t} & =  \frac{h \eta \omega_0}{p_0 R} \,W \label{eq:dphi}
\, , \\
\der{W}{t} & =  \frac{e \hat{V}}{2\pi} \, ( \sin\phi - \sin\phi_{\rm s} ) \: .
\end{align}

The two variables $\phi$ and $W$ are canonical, since these equations of motion can be derived from a Hamiltonian $H(\phi, W, t)$:
\begin{align}
\der{\phi}{t} &= \pder{H}{W}\: , \qquad\der{W}{t} = - \pder{H}{\phi} \: ,
\\
H(\phi,W, t) &= \frac{e \hat{V}}{2\pi} \big[ \cos\phi - \cos\phi_{\rm s} + (\phi-\phi_{\rm s}) \sin\phi_{\rm s} \big] + \frac{1}{2} \frac{h \eta \omega_0}{p_0 R} W^2 \, .
\label{eq:ham}
\end{align}
The basic Hamiltonian shown here reproduces the equations of motion that we found previously.
In more complex cases, the general approach of the Hamiltonian formalism can help us to analyse and understand some fairly
complicated dynamics (multiple harmonics, bunch splitting, \etc).

\begin{figure}[!htb]
\begin{minipage}{\textwidth}
  \centering
  $\vcenter{\hbox{\includegraphics[width=0.59\textwidth]{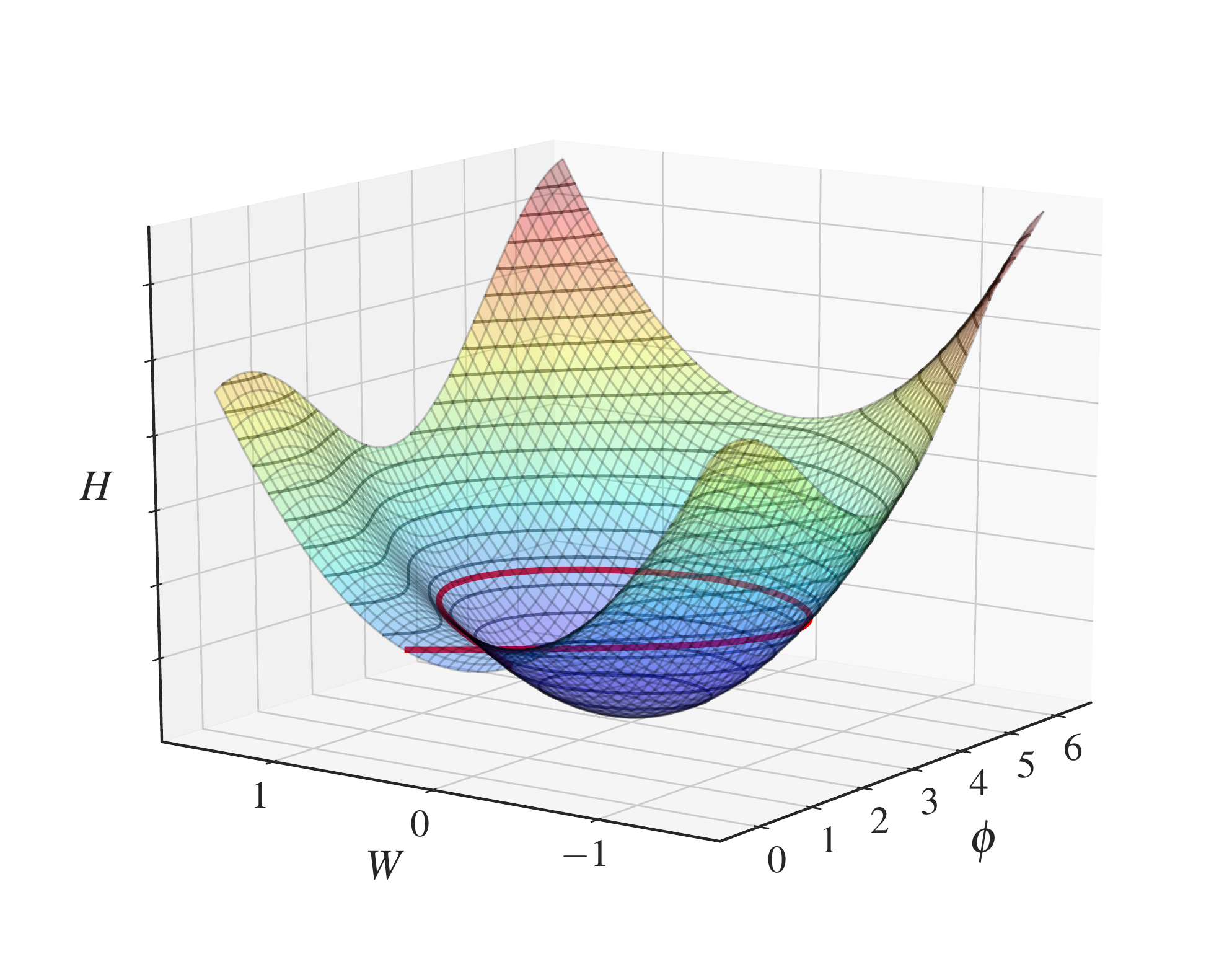}}}$
  \hfill
  $\vcenter{\hbox{\includegraphics[width=0.4\textwidth]{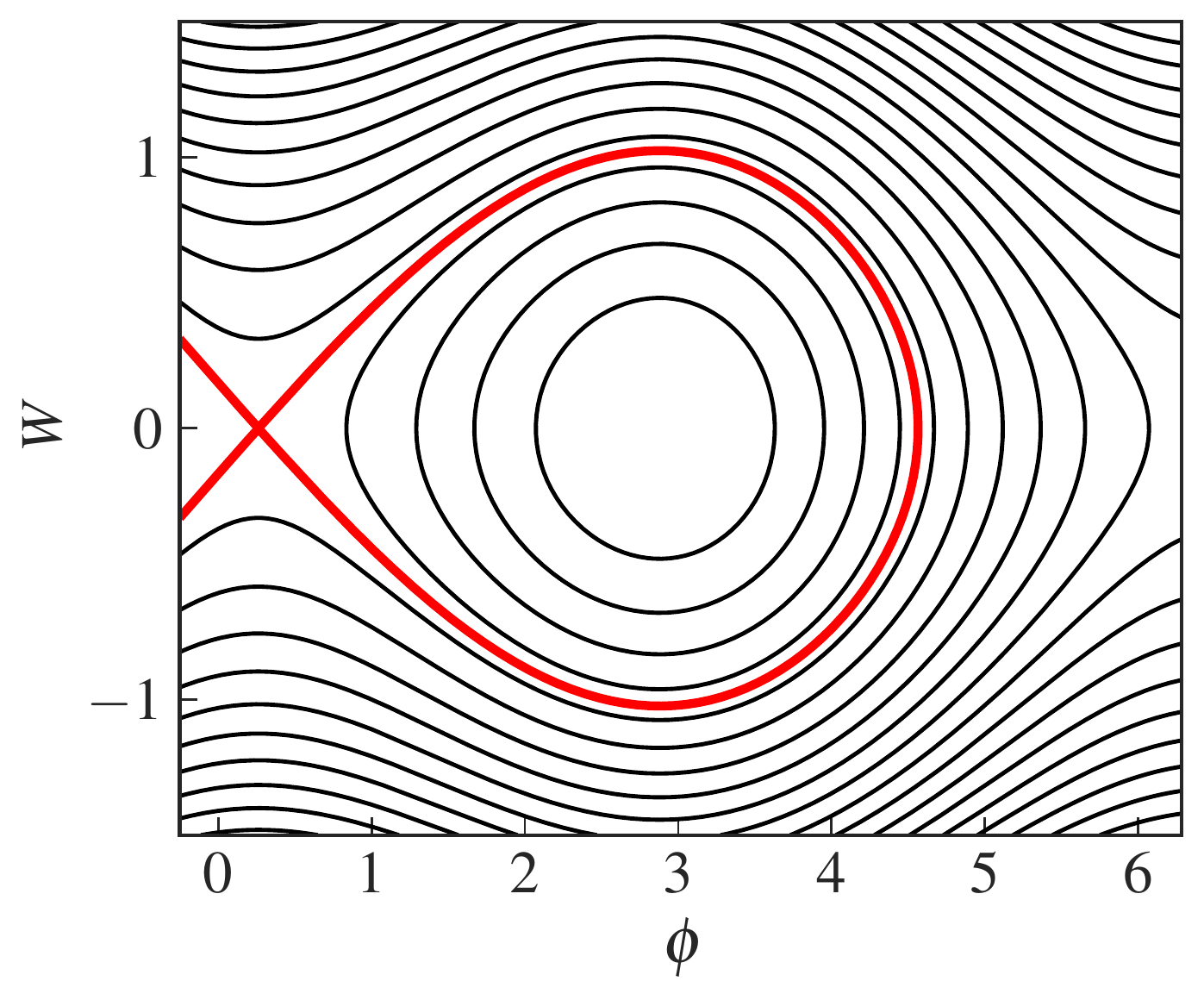}}}$
\end{minipage}
 \caption{\label{fig:phase-space-3d} Example of  a Hamiltonian function $H$ (left) and its projection (right) above transition, for a synchronous phase of $\phi_s=170^\circ$. Equipotential lines are possible phase-space trajectories; the separatrix is shown as a solid red  line.}
\end{figure}

The Hamiltonian $H$ represents the total energy of the system. In fact, if the total energy is conserved,
the contours of constant $H$ are particle trajectories in phase space, as illustrated in Fig.~\ref{fig:phase-space-3d}.

\section*{Bibliography}

J. Le Duff, Proc.\ CAS-CERN Accelerator School: CAS Fifth General Accelerator Physics Course, Jyv\"{a}skyl\"{a}, Finland, 7--18 September 1992 (CERN 94-01). Ed.\ S.\ Turner  (CERN, Geneva, 1994), pp.~253--311, \url{https://doi.org/10.5170/CERN-1994-001}.


\noindent H. Wiedemann, \textit{Particle Accelerator Physics} (Springer, Berlin, 2015),\\ \url{https://doi.org/10.1007/978-3-319-18317-6}.

\noindent K. Wille, \textit{The Physics of Particle Accelerators: An Introduction} (Oxford University Press, Oxford, 2000), \\ \url{https://doi.org/10.1007/978-3-663-11850-3}.


\noindent S. Y. Lee, \textit{Accelerator Physics} (World Scientific, Singapore, 2019),\\
\url{https://doi.org/10.1080/00107514.2019.1641154}.

\noindent A. Wolski, \textit{Beam Dynamics in High Energy Particle Accelerators} (Imperial College Press, London, 2014), \url{https://doi.org/10.1142/p899}.

\end{document}